\documentclass[twocolumn]{openjournal}

\usepackage{newtxtext,newtxmath}
\usepackage[T1]{fontenc}
\usepackage{graphicx}	
\usepackage{amsmath}	

\usepackage{lineno}
\usepackage{xcolor}
\usepackage{color,colortbl}

\usepackage{orcidlink}

\definecolor{linkcolor}{rgb}{0.0,0.3,0.7}
\hypersetup{
    unicode, 
    colorlinks=true,
    linkcolor=linkcolor,
    citecolor=linkcolor,
    filecolor=linkcolor,
    urlcolor=linkcolor,
}

\newcommand{\msun}{\textup{M}_\odot}

\newcommand{\mHcc}{{\rm cm}^{-3}}
\newcommand{\menv}{M_{\rm env}}

\newcommand{\new}{} 
\newcommand{\newnew}{} 

\begin{document}


\title[Unresolved SNe Physics]{The Entangling of Supernova Feedback Impacts with Coarsening Simulation Resolution}
\shorttitle{ Unresolved SNe Physics }
\shortauthors{Zhang et. al.}

\author{Eric Zhang\,\orcidlink{0000-0002-7611-8377}$^{1\star}$}\thanks{$^\star$E-mail: ezhan039@ucr.edu}
\author{Laura V. Sales\,\orcidlink{0000-0002-3790-720X}$^{1}$}
\author{Thales A. Gutcke\,\orcidlink{0000-0001-6179-7701}$^{2\dagger}$}\thanks{$^\dagger$NASA Hubble Fellow.}
\author{Yunwei Deng\,\orcidlink{0000-0002-7478-6427}$^{3}$}
\author{Hui Li\,\orcidlink{0000-0002-1253-2763}$^{3}$}
\author{Rüdiger Pakmor\,\orcidlink{0000-0003-3308-2420}$^{4}$}
\author{Federico Marinacci\,\orcidlink{0000-0003-3816-7028}$^{5,6}$}
\author{Volker Springel\,\orcidlink{0000-0001-5976-4599}$^{4}$}
\author{Mark Vogelsberger\,\orcidlink{0000-0001-8593-7692}$^{7}$}
\author{Paul Torrey\,\orcidlink{0000-0002-5653-0786}$^{8}$}
\author{Boyuan Liu\,\orcidlink{0000-0002-4966-7450}$^{9}$}
\author{Rahul Kannan\,\orcidlink{0000-0001-6092-2187}$^{10}$}
\author{Aaron Smith\,\orcidlink{0000-0002-2838-9033}$^{11}$}
\author{Greg L. Bryan\,\orcidlink{0000-0003-2630-9228}$^{12}$}

\affiliation{$^{1}$Department of Physics and Astronomy, University of California, Riverside, CA 92507, USA}
\affiliation{$^{2}$Institute for Astronomy, University of Hawaii, 2680 Woodlawn Drive, Honolulu, HI 96822, USA}
\affiliation{$^{3}$Department of Astronomy, Tsinghua University, Haidian DS 100084, Beijing, China}
\affiliation{$^{4}$Max-Planck-Institut f\"{u}r Astrophysik, Karl-Schwarzschild-Stra\ss{}e 1, 85740 Garching bei M\"{u}nchen, Germany}
\affiliation{$^{5}$Department of Physics \& Astronomy ``Augusto Righi'', University of Bologna, via Gobetti 93/2, 40129 Bologna, Italy}
\affiliation{$^{6}$INAF, Astrophysics and Space Science Observatory Bologna, Via P. Gobetti 93/3, 40129 Bologna, Italy}
\affiliation{$^{7}$Department of Physics, Kavli Institute for Astrophysics and Space Research, Massachusetts Institute of Technology, Cambridge, MA 02139, USA}
\affiliation{$^{8}$Department of Astronomy, University of Virginia, 530 McCormick Road, Charlottesville, VA 22903, USA}
\affiliation{$^{9}$Universität Heidelberg, Zentrum für Astronomie, Institut für Theoretische Astrophysik, Albert Ueberle Str. 2, D-69120 Heidelberg, Germany}
\affiliation{$^{10}$Department of Physics and Astronomy, York University, 4700 Keele Street, Toronto, ON M3J 1P3, Canada}
\affiliation{$^{11}$Department of Physics, The University of Texas at Dallas, Richardson, TX 75080, USA}
\affiliation{$^{12}$Department of Astronomy, Columbia University, 538 West 120 Street, New York, NY 10027, USA}

\begin{abstract}
It is often understood that supernova (SN) feedback in galaxies is responsible for regulating star formation (SF) and generating gaseous outflows. However, a detailed look at \new{the small-scale effects of SNe} on the interstellar medium (ISM) in simulations shows that \new{the macroscopic processes of SF suppression and outflow generation proceed in distinct channels.} We demonstrate this finding in two independent simulations of isolated dwarf galaxies with very high ($m_{\rm gas}$ $\sim \msun$) numerical resolution, {\small LYRA} and {\small RIGEL}. \new{Our findings suggest that the macroscopic effect of a given SN on the galaxy is best predicted by its immediate local density.} Outflows are \newnew{primarily linked to} SNe in diffuse regions expanding to their cooling radii on large ($\sim$ kpc) scales, while dense SF regions are disrupted in a localized ($\sim$ pc) manner. However, these separate feedback channels are only distinguishable at very high resolutions capable of following mass scales $\lesssim 10^2 \,\msun$. When averaging on coarser scales, ISM densities are greatly mis-estimated, and variations between different SF and SNe-affected regions are severely washed out. It therefore cannot be \textit{self-consistently} determined from coarse-resolution information \textit{alone}, (1) whether a SN tends to contribute to outflows or direct SF suppression, and (2) the rate of SF in a given region. In particular, commonly used parameters in coarse-resolution (subgrid) models, such as the SN cooling radius and SF density threshold, may require more detailed treatments informed by high-resolution studies.
\end{abstract}



\section{Introduction}\label{sec:intro}

Supernovae (SNe) are considered to be one of the most important processes governing the evolution of galaxies. Specifically, they \new{play a crucial role in} (\textit{i}) regulating the star formation (SF) within galaxies, which if controlled by gravitational collapse alone would exceed the observed rates of SF \citep[e.g.,][]{Zuckerman74,WhiteFrenk91,WilliamsMcKee97,Kennicutt98,Evans99SF,KrumholzTan07,Keres09,Evans09Obs,Tasker09SF,Tasker11SF,Dobbs11,Hopkins11}, and (\textit{ii}) generating energetic galactic-scale outflows into the circumgalactic medium (CGM) moving with velocities on the order of hundreds of kilometers per second \citep[e.g.][]{Martin99,Veilleux05,Steidel10,Coil11}. 

The interplay between these processes, however, is not immediately clear. The suppression of star formation by SNe may proceed in many different channels: by slowing the rate at which collapsing gas forms stars, disrupting the star-forming region entirely, removing eligible gas from the galaxy, or preventing gas from ever accreting onto the galaxy. The extent to which the blast wave can drive outflows is also varied; the size to which a SN shell expands to is highly sensitive to its environmental conditions \citep[e.g.,][]{Iffrig15, KimOstriker15, Martizzi15, WalchNaab15, Ohlin19, Hopkins2024}. It is furthermore unclear to what degree the outflows are themselves related to each of the channels by which star formation can be suppressed.

Due to complex interactions between the relevant forces, these processes are commonly studied by numerical simulations, ranging from tall-box simulations of regions of a galactic disk \citep[e.g.,][]{Walch15SILCC,Ostriker22,Sike25}, to entire idealized disks \citep[e.g.,][]{Agertz10,Hopkins11,Kannan20MWRT,Li20Smuggle,Li22SMUGGLE,ZhangZhijie25,IMLADRIS26}, to those of cosmological initial conditions \citep[e.g.,][]{Vogelsberger13,Schaye15EAGLE,NIHAOWang15,Grand17Auriga,The300Cui18, FIRE2,FIRE3, Nelson18TNG, MarvelMunshi19, Agertz20EDGE, Thesan, ThesanZoom}. 

\new{Such simulations include a variety of processes, in addition to SNe, that are also credited with driving outflows and suppressing star formation. These relevant processes include, among others, energy from active galactic nuclei \citep[e.g,][]{Meier99AGN,Birzan04AGN,DeVilliers05AGN,Nemmen07AGN,Komissarov07AGN,Rupke14AGN,Harrison14,RamosAlmeida19AGN,Aravindan23Outflows}, amplification of outflows by cosmic rays \citep[e.g.,][]{Booth13,Salem14,Pakmor16CR,Girichidis18,Rathjen21,Farcy22,PANDORA23}, and “early'' stellar feedback (e.g., photoionization, radiation pressure, OB winds) that can disperse dense star-forming clumps before SNe \citep[e.g.,][]{Heckman00RadPress,Matzner02OB,KrumholzMatzner09OB,VazquezSemadeni10,Walch12,Dale14,Sales14,Rosdahl15PI,Emerick18Rad,Smith21}.}

Modeling SNe in galaxy simulations is achieved in various ways. Direct thermal energy injection of $\sim 10^{51}$ erg into a small number of gas \newnew{parcel(s)} surrounding a star is the simplest energy deposition scheme. \newnew{This approach best reflects first principles in cases where the expansion of this superheated gas is explicitly resolvable, but is} only feasible at the highest mass \citep[$\sim 20 \, \msun$, e.g.,][]{SmithSijackiShen18} and space \citep[$\Delta x \sim 1\,{\rm pc}$, e.g.,][]{KimOstriker15} resolutions. The inability of lower-resolution simulations to simultaneously and explicitly trace the gas’ thermal expansion and radiative cooling is known as the \new{numerical}\footnote{\new{Not to be confused with the classical overcooling problem \citep[e.g.,][]{WhiteRees78,Cole01,Benson03}, whereby efficient cooling in low-mass halos might lead to much more luminous galaxies if feedback is not accounted for.}} overcooling problem \citep[e.g.,][]{Katz92,Vogelsberger20}.

The required resolution is not reached by most simulations of galaxy formation, and so they require ``subgrid'' models that approximate the SN evolution on sub-resolution mass and length scales. This can be done by modified thermal injection \citep[e.g.,][]{Murante10, DallaVecchia12, Chaikin22Numeric, COLIBRE26}, disabling radiative cooling while the SN is expanding \citep[e.g.,][]{Stinson06,Agertz10,Teyssier13}, anticipating and injecting the amount of momentum generated during the thermal expansion \citep[e.g.,][]{Navarro93, Mihos94, Vogelsberger13, Martizzi15, FIRE2,FIRE3}, or the total energy and momentum distribution from multiple clustered SNe \citep[e.g.,][]{Keller14,ElBadry19SB}.

All such models \new{aim to mimic the effects of releasing a given amount of energy under the present conditions of the local interstellar medium (ISM)}. Implicitly, it is assumed that the ISM conditions can be derived from what is available within the simulation. This information is, of course, dependent on the simulation's resolution; coarse-resolution simulations do not carry information about the detailed and multi-phase ISM structure that lies within a single gas \newnew{parcel}. Therefore, the way energy and momentum distributions evolve afterward may drastically vary, even under implementations of the same feedback physics (such as the energy released).

In this paper, we \new{primarily} analyze isolated-disk simulations of similar dwarf galaxies that have been previously presented in the {\small LYRA} \citep[][]{Lyra2021} and {\small RIGEL} \citep[][]{Deng24RIGEL} simulations. The simulations in question are both very high resolution simulations (at target masses of 4 and 1 $\msun$ per cell, respectively), which is \newnew{almost always} enough to explicitly trace the thermal expansion phase of SN remnants. \new{The dominant processes involved in the simulations are self-gravity and hydrodynamics, the formation of stars from gas, the radiative cooling of gas, and the energy released by SNe.} \new{Notably, we have chosen to analyze simulations that do not model radiative feedback from young stars, although it is known to be a vital process in galactic \newnew{gas} dynamics. This allows us to hone in on the effect of energy released by SNe without any confounding energy sources to their local environment.}

In Section~\ref{sec:simulations} we discuss the simulation set-up along with the star formation and stellar feedback procedures in the simulations. In Section~\ref{sec:environment} we describe the effects individual SNe have on their environments under the dominant effects of SNe and radiative cooling, as well as their consequences for the whole galaxy \new{in high-resolution simulations}. In Section~\ref{sec:resolution} we show how resolution impacts all results and how even at moderately low resolution, important aspects of galaxy physics such as outflows cannot be resolved with high fidelity. In Section~\ref{sec:discussion} we discuss the implications of our results, and we provide a summary in Section~\ref{sec:conclusion}.

\section{Simulations}\label{sec:simulations}

We analyze the {\ttfamily Variable\_SN\_energy} simulation from {\small LYRA} \citep[][]{Lyra2021} and the {\ttfamily0.02${\rm Z}_{\odot}$/noRT} simulation from {\small RIGEL} \citep[][]{Deng24RIGEL}. The data available to us consists of {\small LYRA} snapshots every 1 Myr and {\small RIGEL} snapshots every 10 Myr, with total time 1 Gyr. Both galaxies are simulated using the moving mesh hydrodynamics code {\small AREPO} \citep[][]{AREPO2010,Weinberger20AREPO} \new{with self-gravity}, and commonly-used initial conditions from \citet{Hu17}. 

These initial conditions are similar but not identical: they consist of a low baryon-to-halo mass ratio (0.3 \% baryon fraction) dwarf with a dark matter halo of mass $M_{\rm vir}=2\times10^{10}\,\msun$ with virial radius 44 kpc and a concentration $c_{\rm halo}=10$. Of the baryons, 1/3 of the mass is allocated to a pre-existing stellar disk, following an exponential profile with a scale radius of 0.73 kpc and a scale height of 0.35 kpc; the remaining 2/3 is allocated the gas disk. The gas disk also follows an exponential profile with the same radius and with height such that the disk is initially in hydrostatic equilibrium. The initial stellar component of both simulations is distributed in the same way, but the gas disk is different: in {\small LYRA} its scale radius is the same as the stellar disk's, and it is first relaxed with stochastic energy injection for 200 Myr; in {\small RIGEL} it is twice this length, and first relaxed for 500 Myr. \new{We consider the end of the relaxation period of both simulations to be $t=0$.} Furthermore, the initial gas metallicity of {\small LYRA} is 0.01 $Z_{\odot}$; in {\small RIGEL} it is 0.02 $Z_{\odot}$. The simulations contain no significant CGM, \new{and thus do not model the accretion of additional gas onto the disk from the CGM}. {\small LYRA} and {\small RIGEL} have very high target resolutions of 4 and 1 $\msun$ respectively. {\small AREPO}'s mesh refinement scheme ensures that the mass of any given gas cell is within a factor of 2 from the target mass. \new{The gravitational softening length is 0.5 pc in {\small LYRA}, and from 0.004 to 0.05 pc in {\small RIGEL}, determined adaptively.}

Beyond standard hydrodynamics \new{and self-gravity}, each simulation has its own methods of modeling stellar processes within the galaxy. Chief among these processes are star formation and supernova feedback, which will be detailed in the following subsections. While not a source of stellar feedback, radiative cooling of baryons is also included in the suite of additional physics. \new{As will be detailed in the following sections, the numerical parameters modulating these physics are likewise different between simulations, producing nontrivial differences in the resulting dynamics of the galaxies.} Radiative feedback from young stars, while known to be a crucial process in galaxy evolution in general, is \textit{not} included in the simulations studied in this paper.

\subsection{Star Formation}\label{subsec:sfrmodel}

Stars are modeled using collisionless particles, which are created by transforming eligible gas cells. In both simulations, gas is stochastically converted into star particles at a rate consistent with the Schmidt law
\begin{equation}\label{eq:cellSFR}
    \dot{M}_{\rm SF} = \epsilon_{\rm SF} \frac{M_{\rm cell}}{t_{\rm ff}},
\end{equation}
where $\epsilon_{\rm SF}$ a tunable parameter representing the efficiency of star formation \new{per free-fall time}, $M_{\rm cell}$ is the mass of the gas cell and $t_{\rm ff}$ is the \newnew{free-fall} timescale of the gas in the cell, given by:
\begin{equation}\label{eq:tdyn}
    t_{\rm ff} = \sqrt{\frac{3 \pi}{32 G \rho_{\rm cell}}},
\end{equation}
where $G$ is the gravitational constant and $\rho_{\rm cell}$ is the gas cell density. Furthermore, the simulations also employ a density and temperature threshold for star formation, where stars can only be formed from dense and cold gas. In particular, only gas cells with density $\rho_{\rm cell} > \rho_{\rm th}$ and temperature $T<T_{\rm th}$ are eligible for star formation. In {\small LYRA} the efficiency of star formation $\epsilon_{\rm SF}$ is set to 0.02, and the density threshold $\rho_{\rm th}$ is set to $10^3$ $\mHcc$; in {\small RIGEL} the efficiency is set to 1, and the threshold is set to $10^4$ $\mHcc$. The temperature threshold is $T_{\rm th}=100 \,{\rm K}$ in both simulations. 

Tuning such parameters has been shown to meaningfully affect galaxy evolution. Changing the efficiency parameter $\epsilon_{\rm SF}$ controls the degree of SF clustering \citep[e.g.,][]{Hislop22}, and in some studies can change the overall SF level by a factor of $\sim 2$ \citep[e.g.,][]{Agertz13Energy,Zhang24}. \newnew{The temperature threshold may also affect the overall SF level in general, although this sensitivity is primarily a concern either at lower resolutions or in the presence of radiative feedback \citep[e.g.,][]{Rey25}, which is not included in this study.} In addition, a high SF threshold $\rho_{\rm th}$ allows for more cold gas to build up before being disrupted, leading to bulk gas flows that can eventually flatten dark matter cusps \citep[e.g.,][]{PontzenGovernato12, BenitezLlambay19} and change the global gravitational potential of the galaxy.

\new{One way to view these dynamics is via estimating the timescale at which SF-eligible gas is converted into stars, via $t_{\rm SF}=t_{\rm ff}/\epsilon_{\rm SF}$ where the \newnew{free-fall} timescale is calculated at the SF density threshold $\rho_{\rm th}$ of the simulation. In {\small LYRA} this timescale is 80 Myr, but in {\small RIGEL}, it is 0.5 Myr. Thus, via a combination of choices of $\rho_{\rm th}$ and $\epsilon_{\rm SF}$, cold dense gas in {\small LYRA} is allowed to cluster for far longer before either forming stars or being dispersed by feedback.}

{\small RIGEL} further imposes a ``virial parameter condition'' \citep[e.g.,][]{Hopkins13Vir,Semenov17Vir} requiring that the gas be locally gravitationally bound, and also that the Jeans mass must also be resolved, in order to form stars. The Jeans length at the density and temperature threshold is $\sim 3 \,{\rm pc}$ for {\small LYRA} and $\sim 1 \, {\rm pc}$ for {\small RIGEL}, which is larger than the softening lengths of their respective simulations, so that the softening does not cause artificial fragmentation.

The masses of stars at their birth are drawn from an initial mass function (IMF); {\small LYRA} uses a \citet{KroupaIMF} IMF sampled from 0.01 to 100 $\msun$, whereas {\small RIGEL} uses a \citet{ChabrierIMF} IMF sampled from 0.1 to 100 $\msun$. The mass resolution is sufficiently high so that individual massive stars and their properties, such as its time of death and resulting SN yield, are resolved. The IMF in both simulations does sample stellar masses below their minimum resolution; in such cases, individual stars are considered unresolved, and the star particle represents an aggregate stellar population. \newnew{In {\small LYRA} the mass of these unresolved star particles is enforced to be the minimum gas cell mass of 4 $\msun$; in {\small RIGEL}, these star particles simply inherit the mass of their progenitor gas cells.}

\subsection{Supernova Feedback}\label{subsec:snfeed}

In both simulations, massive stars above a threshold of 8 $\msun$ explode as Type II supernovae, and mass, metals, and energy are deposited back into the ISM at the time of the star's death. In the {\small RIGEL} model, white dwarfs may also explode as Type Ia supernovae according to their respective stellar evolution models. 

Both simulations have sufficient resolution to model both individual massive stars and their resulting SNe, and furthermore, for these SNe to be modeled as point thermal energy injections of $\sim 10^{51}$ erg into a single gas cell. \newnew{Since their resolution is sufficiently high at 4 and 1 $\msun$ for {\small LYRA} and {\small RIGEL} respectively, the numerical overcooling problem is not a concern and} the expansion of the superheated gas into the ISM can be studied explicitly.

{\small RIGEL} simply injects $1.0 \times 10^{51}$ erg per event, for every star between 8 and 100 $\msun$. {\small LYRA} instead employs a variable SN energy scheme informed by models of stellar cores from \citet{Sukhbold16}. In this scheme, based on the mass of the progenitor star, the SN injects energy on the scale of $10^{51}$ erg, but varying from 0.1 up to $1.8 \times 10^{51}$ erg. \newnew{The resulting average energy injected is $0.44 \times 10^{51}$ erg per massive star and $0.46 \times 10^{51}$ erg per 100 $\msun$ of total stellar mass \citep[see Table 2 in ][]{Lyra2021}.} The energy injected is not a monotonic function of the progenitor mass, and for some masses no energy will be injected at all, including almost all stars above 30 $\msun$ \citep[see Fig. 2 in ][]{Lyra2021}.

While the energy is always released into the nearest gas cell, the mass and metal release scheme differs between the simulations. For {\small LYRA}, mass and metals are released into the nearest gas cell alongside the energy, but for {\small RIGEL}, they are released into the nearest 32 gas cells, which is different from the energy injection scheme. The metals will then diffuse into the galactic gas according to the hydrodynamics solver. Mass and metals are returned to the ISM regardless of whether any energy is released, such as in {\small LYRA} cases where the variable SN energy is zero.

\subsection{Gas Cooling}

For high-temperature gas, cooling is done using {\small CLOUDY} \citep[][]{Ferland98,Ferland17} tables in all simulations. For low-temperature ($\lesssim 10^4$ ${\rm K}$) gas, the method varies; {\small LYRA} still uses tabulated {\small CLOUDY} values, as discussed in Sec. 2.2 of \citet{Lyra2021};  {\small RIGEL} re-derives the low-temperature cooling rate via equilibrium abundances of C and O; see the discussion in Sec. 2.2 of \citet{Deng24RIGEL}.

\section{Effect on Local Environments}\label{sec:environment}

\begin{figure}
\centering
\includegraphics[width=0.98\columnwidth]{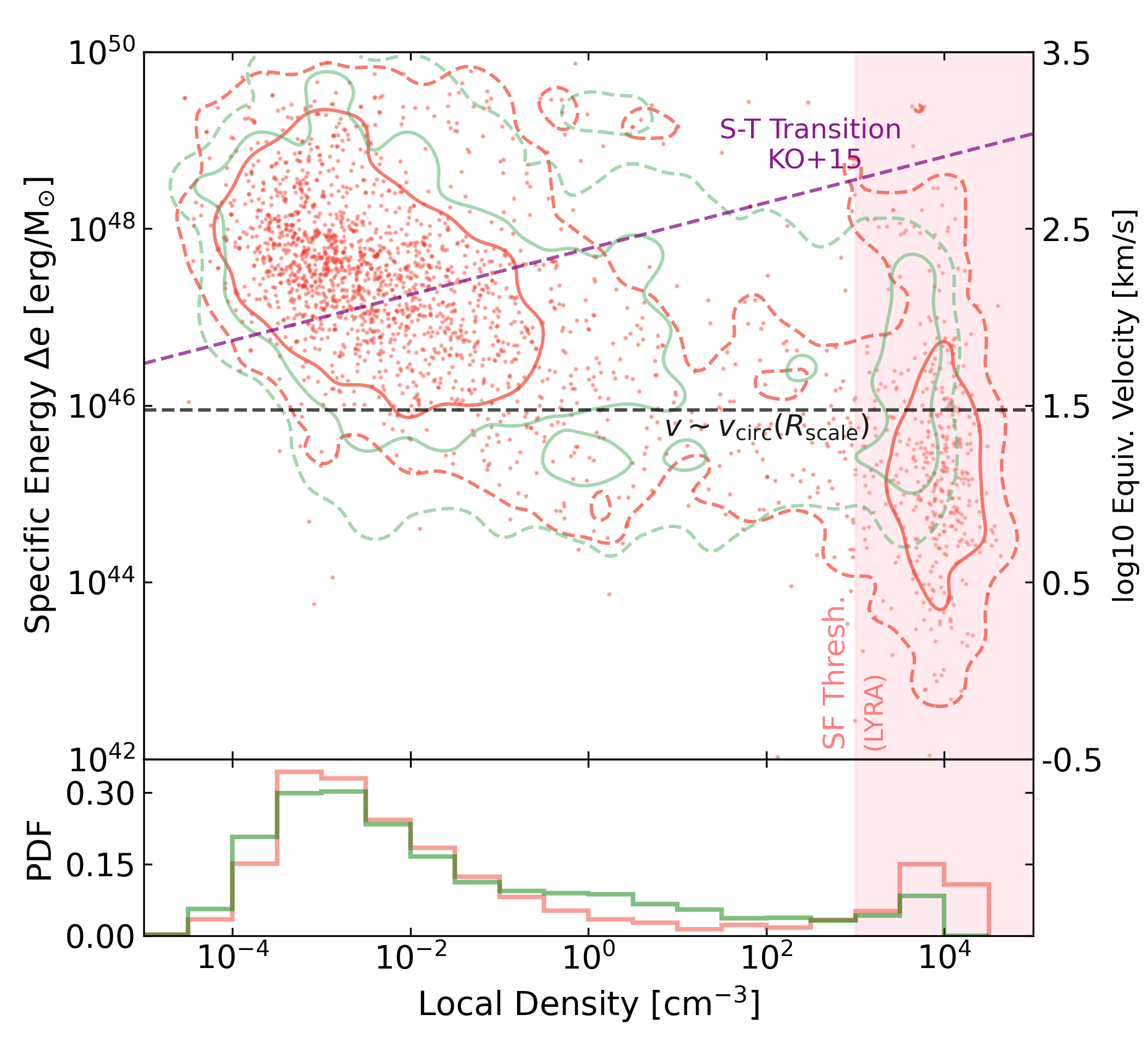}
    \caption{\textit{Bimodal Distribution of Local Environments to SNe.} \textbf{Top:} Change in specific energy $\Delta e$ vs the local density of each SN as measured using the nearest $\menv=20\, \msun$ to the SN, in the snapshots immediately before and after the SN. $\Delta e$ need not be only due to the SN in question; nearby SNe and radiative cooling will also affect its value. Each SN in LYRA is a red point, alongside 68\% and 95\% contours of their distribution in solid and dashed red lines. The same contours from RIGEL are shown in green. Equivalent velocities to each $\Delta e$ via Eq.~(\ref{eq:velocity}) are shown in the right-hand tick marks. In the gray dashed line, we plot the $\Delta e$ level corresponding to the circular velocity at the initial LYRA scale radius of 0.73 kpc. In the purple dashed line, we plot, as a function of density, the analytically predicted specific energy of a SN blastwave at the end of the Sedov-Taylor phase \citep[e.g.,][]{KimOstriker15}. The red shaded region shows the gas densities where star formation is possible in LYRA. \textbf{Bottom:} Histogram of local densities to SNe, corresponding to the $x$-scatter in the top panel. Additionally includes SNe with $\Delta e \leq 0$, which are not shown in the top panel.}
    \label{fig:BasicEnergy}
\end{figure}

\subsection{Analysis of Individual SNe}\label{subsec:indivSNe}

First, we identify the time and location of each SN event in the two simulations. We then measure the properties of the local environment in the simulation snapshots before and after the SN event, where we denote the snapshot separation time by $\Delta t$. Based on the snapshots which we have access to, $\Delta t = 1\,{\rm Myr}$ for {\small LYRA} and $\Delta t=10\,{\rm Myr}$ for {\small RIGEL}, giving a higher time resolution for our study of {\small LYRA} \newnew{than of {\small RIGEL}.} Furthermore, for {\small LYRA} specifically, we ignore SNe that release zero energy due to the variable energy scheme.

The local environment is considered to be the gas cells that comprise the nearest $M_{\rm env} = 20\,\msun$ of gas to the star particle on any given snapshot. We characterize the environment in which the SN goes off by \newnew{its} density $\rho$ and \newnew{its} change in specific energy $\Delta e$, including both kinetic and thermal, between the two snapshots. Specifically,
\begin{equation}\label{eq:rho}
    \rho = \frac{M_{\rm env}}{V_{\rm env}}
\end{equation}
where $V_{\rm env}$ is the smallest spherical volume containing all cells comprising $\menv$, and
\begin{equation}\label{eq:de}
    \Delta e = \frac{E_{\rm after} - E_{\rm before}}{\menv}
\end{equation}
where $E_{\rm after}$ and $E_{\rm before}$ are the total kinetic and thermal energies of all cells comprising $\menv$. 

In Fig. \ref{fig:BasicEnergy}, we plot $\Delta e$ against $\rho$ for each SN in {\small LYRA} (red), along with the 68\% and 95\% contours of their distribution. The contours for ${}${\small RIGEL} are shown as well (green), and are consistent with the {\small LYRA} distribution. The data indicates clearly two groups or ``modes"\newnew{:} a population of supernovae going off in low densities $n \sim 10^{-4}-10^{-1}$ cm$^{-3}$, that are associated with a large change in the local energy $\Delta e$, and a population of supernovae in high densities $n \sim 10^{4}$ cm$^{-3}$, associated with a smaller $\Delta e$ change. These two populations, well-separated in density, have been noted previously by various studies \citep[e.g.,][]{Lyra2021,Hislop22}. Due to very efficient cooling at high densities, the typical difference in $\Delta e$ between the two modes is 2 to 3 orders of magnitude, but it is possible for $\Delta e$ to differ by as many as 7 orders of magnitude.

The specific energy change $\Delta e$ can be understood as a velocity, by calculating the equivalent velocity if all of $\Delta e$ were kinetic, via
\begin{equation}\label{eq:velocity}
    v_{\rm equiv.} = \sqrt{2 \Delta e}.
\end{equation}
If all of the energy within a heated gas region were to be converted to kinetic, the gas would be accelerated to this velocity. Using this conversion, we plot the circular velocity ($v_c = 30\,{\rm km/s}$) at the galaxy's scale radius ($0.73\,{\rm kpc}$) alongside the distribution of $\Delta e$ in Fig.~\ref{fig:BasicEnergy}. This approximates a \new{conservative} threshold at which gas is \textit{eligible} to escape the disk, as gas pushed upward (or downward) with $v_{\rm equiv.}$ from the disk can at most enter a circular orbit about the galactic center. Gas must have \textit{at least} this much energy to escape into the CGM, while gas with energy below this threshold must remain confined to the disk. The majority of SNe in the low-density (high-density) mode have $\Delta e$ lying above (below) this threshold. 

We stress that this threshold is intended as a conservative \textit{lower limit}, where gas with energy below this level is certain to remain in the disk, but gas with energy above this level is still not guaranteed to become an outflow. SNe lying above the threshold need not convert all their energy into kinetic form \newnew{and} may still lose significant energy before breaking into an outflow (e.g., via additional radiative cooling or ram pressure). \newnew{Furthermore, it is sufficient within the context of this work for gas to merely break into the CGM without necessarily escaping the halo. Thus, we do not set this threshold higher, e.g., at the escape velocity $v_{\rm esc.}=\sqrt{2}v_c$.}

\begin{deluxetable*}{lcccccr}
\tabletypesize{\footnotesize}
\tablecaption{Supernova Statistics}
\tablehead{
SN Mode & $\Delta e>v_c^2/2$ & $0 < \Delta e<v_c^2/2$ & $\Delta e<0$ & $\Delta e>e_{\rm ST}$ & ${\rm SFR}=0$ Bef. & ${\rm SFR}=0$ Aft.
}
\startdata
\label{tab:snstats}
    Low-Density & 74.7\%  (\textit{56.5})& 4.94\% (\textit{8.56})& 20.3\% (\textit{34.9})& 50.6\% (\textit{35.8})&99.7\% (\textit{98.2})&99.8\% (\textit{96.7})\\
    High-Density & 28.6\% (\textit{62.0})& 68.9\% (\textit{36.3})& 2.56\% (\textit{1.70})& 2.56\% (\textit{5.02})&23.8\% (\textit{92.8})&98.4\% (\textit{92.7})
\enddata
\tablecomments{Percentages of low- and high-density SNe shown in Fig.~\ref{fig:BasicEnergy} that fall within relevant bins of $\Delta e$ for {\small LYRA} (with {\small RIGEL} in parentheses). A low- (high-) density SN is defined as one occurring in a density less (greater) than $10^1 \, \mHcc$. We also summarize the effect of these SNe on local star formation.}
\end{deluxetable*}

We assume that the bulk of changes to the energy are caused by either SNe or cooling. It is possible, though, for the local $\menv$ to a given SN to be affected by a \textit{different but nearby} SN, due to SN clustering. Clustered SN feedback can be especially common in simulations that do not include early radiative feedback \citep[e.g.,][]{Smith21}, such as those studied here. In these cases, the main contributor to the $\Delta e$ of a given SN may not be due to the SN itself. \new{Given that for regions of equal mass, high-density regions are much smaller than low-density regions, it is common for high-density regions to overlap with low-density regions. This may cause high-density SNe to have higher $\Delta e$'s compared to what \newnew{they} would \newnew{have} in isolation. Even so, the distribution of $\Delta e$ in high-density SNe is still typically lower than the low-density distribution.}

In cases where cooling is dominant instead of the energy released by SNe, it is also possible for $\Delta e$ to be negative between snapshots. Cooling is especially efficient at high (>$10^4$ K) temperatures, so that when SNe go off in media that was already heated and dispersed by a recent nearby SN, the region may cool over the snapshot interval $\Delta t$ to a temperature lower than what it was before the event. Such SNe \new{with $\Delta e < 0$} almost invariably belong to the low-density mode. \new{We include these SNe in the density histogram} in the bottom panel of Fig. \ref{fig:BasicEnergy}, but we do not show them in the scatter in the top panel \new{as their negative $\Delta e$ is incompatible with the log scale.}

We highlight the cases \newnew{in the two preceding paragraphs}, where the $\Delta e$ associated with a given SN is primarily affected by other nearby SNe or by cooling, to emphasize that each $\Delta e$ need not be the result of each SN in isolation. Indeed, as we are studying the impact of SNe in realistic galaxy settings, the media surrounding each SN necessarily includes all relevant processes, including those not due to the SN in question.

The purple dashed line in Fig.~\ref{fig:BasicEnergy} shows the specific energy as predicted by the \citet{KimOstriker15} analytic solution (for a blastwave of $1.0 \times 10^{51}$ erg), at the end of the energy-conserving (or Sedov-Taylor) phase,
\begin{equation}\label{eq:specE_ST}
    e_{\rm ST}(\rho) = \frac{10^{51}\,{\rm erg}}{1680 \,\msun \cdot (\rho / \mHcc)^{-0.26} }
\end{equation}
where the denominator is the swept-up mass given by Eq.~11 in \citet{KimOstriker15}. In the analytic solution, the total energy of the SN remnant is roughly constant until the end of this phase, at which point radiative losses become important and the total energy declines. \newnew{A significant proportion of low-density SNe environments have $\Delta e$ lying above this reference threshold.}



\begin{figure}[t]
\centering
\includegraphics[width=0.98\columnwidth]{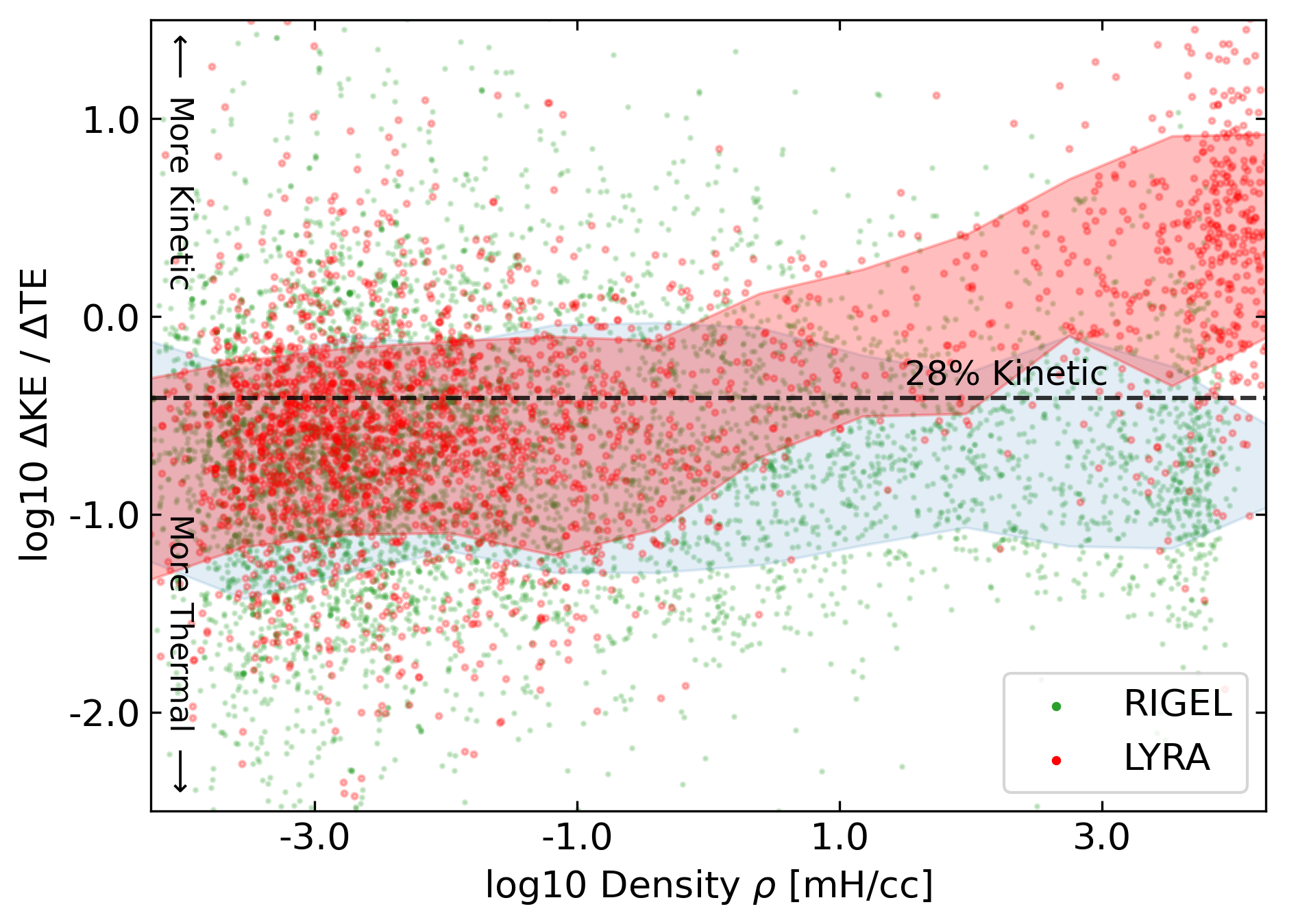}
    \caption{\textit{Ratio of changes in kinetic to thermal energy.} Ratio of the change over the interval $\Delta t$ in kinetic energy $\Delta {\rm KE}$ to the change in thermal energy $\Delta {\rm TE}$, against the local density $\rho$, in the nearest $\menv=20\,\msun$ to each SN. Each point corresponds to one SN, with only SNe that have $\Delta e>0$ considered. The shaded regions correspond to the 1-$\sigma$ range of this ratio at each density. We show LYRA in red and RIGEL in green. The ratio predicted by the analytic solution during the Sedov-Taylor phase is 28\% kinetic energy to 72\% thermal energy, which we show as a gray dashed line. The low-density SN mode tends to be centered about this ratio. The high-density mode leans towards higher kinetic energy in LYRA, while in RIGEL it continues to be centered about this ratio.}
    \label{fig:ratio}
\end{figure}

The analytic solution for the energy-conserving phase of an idealized SN predicts a constant ratio of 28\% kinetic energy to 72\% thermal energy during the phase. Fig.~\ref{fig:ratio} compares this value with the energy ratio $\Delta {\rm KE} /\Delta {\rm TE}$ comprising the energy change $\Delta e$ for each {\small LYRA} and {\small RIGEL} SN. Environments in the low-density mode for both simulations tend to be thermal-favored, and the median energy ratio at such densities is close to the analytic value. For {\small LYRA} in particular, the high-density mode instead favors kinetic energy, indicating a transition from the Sedov-Taylor to the snowplow phase, in which radiative losses have cooled away most of the thermal energy so that the remaining energy is primarily carried by momentum in the swept-up shell. The environments surrounding high-density SNe in {\small RIGEL}, meanwhile, have similar energy ratios as those of low-density SNe and the analytic ratio, though we will show that this may be an effect of the snapshot cadence $\Delta t$. 

Having established that two modes of SNe exist, we define a low-density (high-density) SN as one going off in a local $\rho$ of less than (greater than) $10^{1}\,{\rm cm^{-3}}$. This boundary is well in between the peaks in density and thus separates the two modes well. In Table~\ref{tab:snstats} we summarize the percentages of each SN mode falling within the energy bins $\Delta e \geq v_c^2 /2$, $0 < \Delta e < v_c^2/2$, and $\Delta e \leq 0$. We include values for both {\small LYRA} and {\small RIGEL}, with values for {\small RIGEL} in parentheses. We also consider the percentage of SNe with $\Delta e > e_{\rm ST}$, as given by Eq.~\ref{eq:specE_ST} and the purple line in Fig.~\ref{fig:BasicEnergy}. 

In addition to changes in the local energy, we also consider the star formation rate of the local $\menv=20\,\msun$ region, defined as the sum of the cell-wise $\dot{M}_{\rm SF}$ as defined via Eq.~(\ref{eq:cellSFR}). In most of the considered regions, this value is zero, and so we tabulate only whether the local SFR is zero or nonzero. \newnew{Like the method used to find $\Delta e$ in Fig.~\ref{fig:BasicEnergy}, the instantaneous SFRs are tabulated in the snapshots immediately before and after the SN, separated by the output time interval $\Delta t$.}

These statistics for low- and high-density SNe are summarized in the last 2 columns of Table~\ref{tab:snstats}. For {\small LYRA}, the vast majority of low-density SN environments are not star-forming at all, both before (99.7\%) and after (99.8\%) the event. This is expected as the ISM densities in these regions ($< 10^1\,\mHcc$) are far below the cell-wise SF threshold. The majority of high-density SN environments, meanwhile, are star-forming before the event (23.8\% non star-forming, or 76.2\% star-forming), but still overwhelmingly non star-forming after the SN (98.4\% non star-forming or only 1.6\% star-forming).

The time resolution of {\small RIGEL} is too coarse to accurately sample changes in star forming regions. This can best be understood by comparing the simulation's snapshot interval to its star formation timescale \new{as discussed in Sec. \ref{subsec:sfrmodel}. {\small RIGEL}'s star formation timescale of 0.5 Myr is much shorter than its $\Delta t$ of 10 Myr, so we miss the evolution of star-forming regions; meanwhile, {\small LYRA}'s star formation timescale of 80 Myr is much longer than its $\Delta t$ of 1 Myr, so that it does not suffer from the same problem.} Thus, while we quote the same statistics on SF regions in Table \ref{tab:snstats} \new{for {\small RIGEL}}, we do not derive any meaningful implications from them; though it is still noteworthy that $> 90\%$ of all {\small RIGEL} SN environments have zero SFR post-SN, \new{showing that SNe are indeed effective at dispersing star-forming clumps as expected.} Instead, we will rely on the {\small LYRA} data to study SF suppression.

\begin{figure}
\centering
\includegraphics[width=0.95\columnwidth]{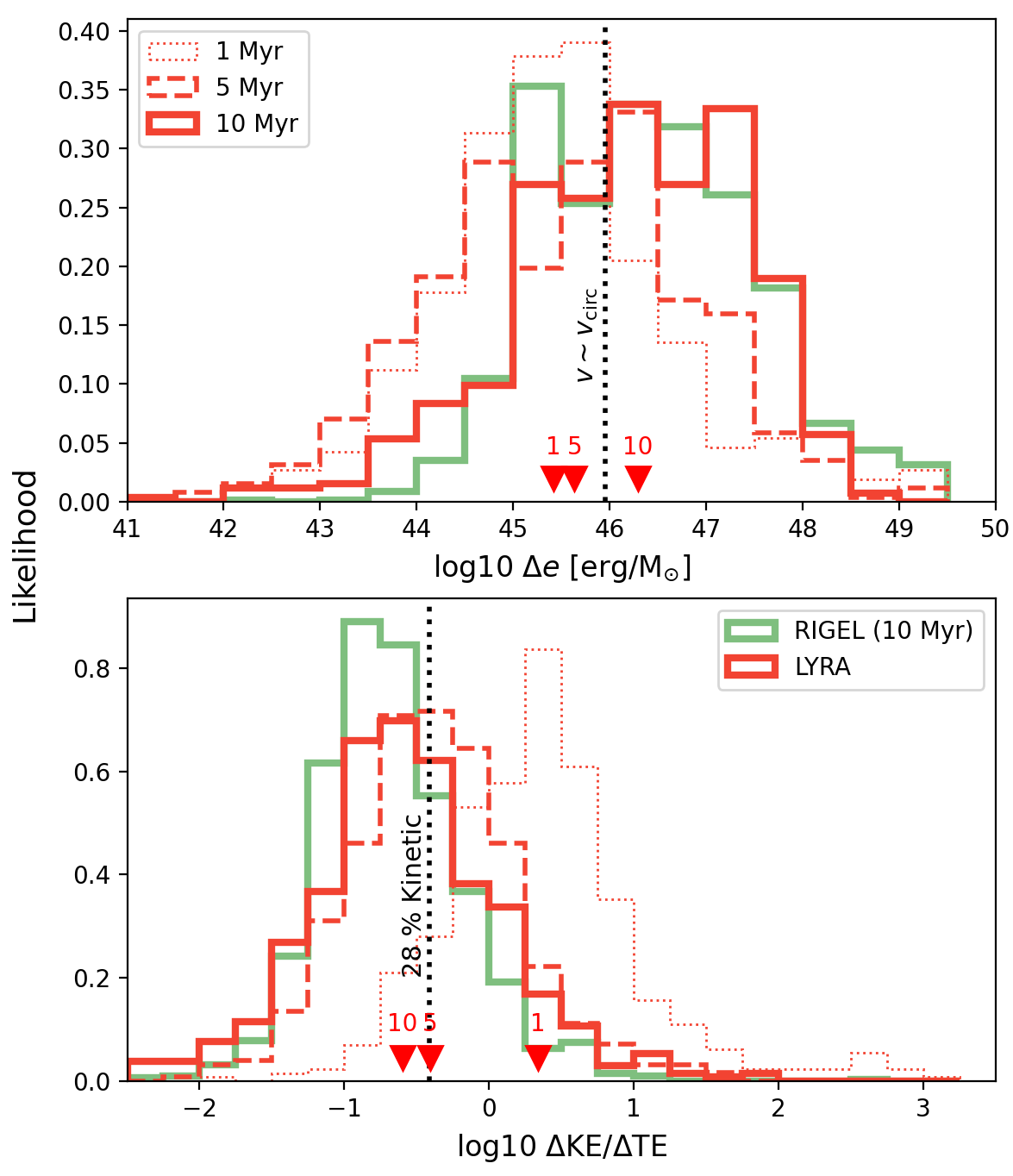}
    \caption{\textit{Dependence of properties on $\Delta t$.} The distribution of $\Delta e$ (top, as per Fig. \ref{fig:BasicEnergy}) and the kinetic-to-thermal injection ratio (bottom, as per Fig. \ref{fig:ratio}) among \textit{high-density SNe} at various $\Delta t$. LYRA results (red) are shown at the base snapshot cadence $\Delta t=1 \, {\rm Myr}$ (dotted) and at the coarser cadences of 5 (dashed) and 10 (solid) Myr. RIGEL results (green) are shown at $\Delta t=10\, {\rm Myr}$.}
    \label{fig:dt}
\end{figure}

\new{For low-density SNe in {\small LYRA}, we find that the distributions of $\rho-\Delta e$ and the kinetic-to-thermal injection ratio are robust for longer $\Delta t$ up to 10 Myr, which is used by {\small RIGEL}. The distributions among \textit{high-density} {\small LYRA} SNe are less robust, and so we display them at various $\Delta t$ values in Fig.~\ref{fig:dt}. The $\Delta e$ distribution is virtually unchanged from $\Delta t=$1 to 5 Myr, but shifts upward for $\Delta t=$10 Myr so that the median $\Delta e$ increases by 0.88 dex. The proportion of high-density SNe beneath the circular velocity threshold decreases from 69\% to 38\% as a result, but we remind the reader that the threshold is conservative. The median ratio $\Delta {\rm KE}/\Delta{\rm TE}$, shifts from a kinetic-favored value of $10^{0.33}$ at $\Delta t = 1\,{\rm Myr}$ to $10^{-0.60}$ at 10 Myr, more consistent with the energy-conserving phase and low-density SNe. At $\Delta t=10\,{\rm Myr}$, both {\small LYRA} distributions are consistent with {\small RIGEL}.}

Only $\Delta t$ longer than 1 Myr are represented by Fig.~\ref{fig:dt}; we do not expect the distribution to be robust for much shorter intervals, since for $\Delta t$ shorter than the cooling timescale at high densities, the high-density mode would not radiate most of its SN energy away.

\subsection{Outflows}\label{subsec:outflows}

In this section we study energy outflows, focusing on whether energy is efficiently deposited into the CGM. We note that none of the outflows in question are powerful enough to reach the virial radius of 44 kpc; outflows in this context are only those that break above and below the disk to reach the CGM.

\begin{figure}[t]
\centering
\includegraphics[width=0.98\columnwidth]{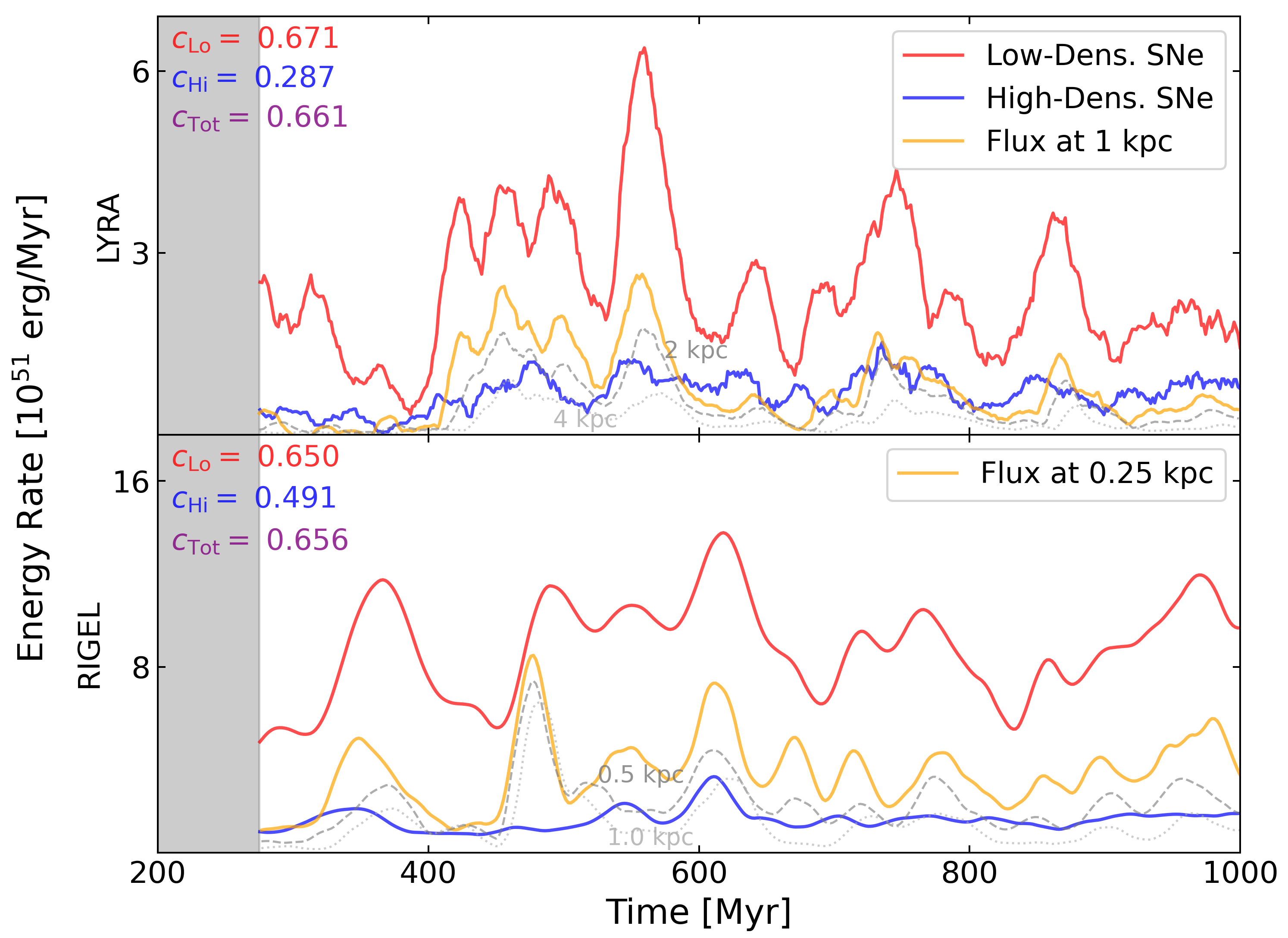}
    \caption{\textit{Low-Density SNe are correlated with energy flows.} SN energy injection rate by low-local density (red) and high-local density (blue) SNe, compared to the energy outflow (orange), as a function of time in LYRA (top, at 1 kpc) and RIGEL (bottom, at 0.25 kpc). The low-density SNe are more temporally correlated to the energy flux than the high-density SNe. Shown in gray are the LYRA energy fluxes at 2 kpc (dashed) and 4 kpc (dotted) for comparison (0.5 and 1.0 kpc in RIGEL). The cross-correlation $c$ of the energy injections from the low- and high-density SNe with the energy flux are printed in the top left corner with their corresponding color. The cross-correlation of all (both low- and high-density together) SNe with the energy flux is also printed (purple).}
    \label{fig:temporal}
\end{figure}

Fig. \ref{fig:temporal} shows the time evolution of the total energy deposited by low-density SNe (red) and high-density SNe (blue). As we have defined in Sec.~\ref{subsec:indivSNe}, a SN is considered to have gone off in a low (high) density environment if the local ISM density is less (greater) than $10$ $\mHcc$. Shown in orange are the energy flux rates calculated from slabs, of thickness $L=50$ pc, at a height of $\pm 1$ kpc in {\small LYRA} and $\pm 0.25$ kpc in {\small RIGEL} above and below the disk, via the equation \citep[][]{Ogilvie16Fluids}: 
\begin{equation}\label{eq:energyoutflow}
    \dot{E}_{\rm out} =
    \frac{1}{L}
    \int
    {\rm d}z
    \int_0^{ R_{\rm max} } {\rm d}A 
    \left(
    \frac{1}{2} \rho v_z^2 + \varepsilon + P
    \right) v_z \, ,
\end{equation}
where $\rho$ is the mass density, $v_z$ is the $z$-velocity directed away from the disk, $\varepsilon$ is the thermal energy density, and $P$ is the pressure. To reduce sensitivity to initial blowouts, we use only the final 750 Myr of each simulation, \new{ignoring the first 250 Myr from $t=0$}.

In both simulations, we see from the alignment of the peak positions, that the red curve representing the low-density SNe are temporally correlated with the energy flow rate. \newnew{Contrarily,} the blue curve representing high-density SNe is not as well-correlated to the outflow history. Relative to the energy released by SNe, the {\small RIGEL} energy flows do not penetrate out of the disk as efficiently as in {\small LYRA}. As such, we measure the {\small RIGEL} flows at 0.25 kpc instead of 1.0 kpc in {\small LYRA}, noting that the peaks in the {\small RIGEL} energy flow at 0.25 kpc visually align with both those of the low-density SNe and the energy flow at the further height of 1.0 kpc.

To formally quantify the association between the outflow rate $\dot{E}_{\rm out}$ and the energy injection rate by SNe $\dot{E}_{\rm SN}$ (referring to low-density, high-density, or all SNe), we define a cross-correlation coefficient
\begin{equation}\label{eq:crosscorr}
    c = {\rm max}_{\tau}
    \langle
    \dot{E}_{\rm SN}, \dot{E}_{\rm out}
    \rangle^2 (\tau)
\end{equation}
where we take a maximum over the lag parameter $\tau$ because the best correlation may occur when one time series is shifted by some time interval $\tau$. Here, $\langle f,g\rangle (\tau)$ is defined analogously to the usual Pearson correlation coefficient, via
\begin{equation}
    \langle
    f,g
    \rangle (\tau) =
    \int \bar{f}(t-\tau) \bar{g}(t)\,{\rm d}t
\end{equation}
where $\bar{f}$, $\bar{g}$ denote the normalized versions of the functions $f$ and $g$, defined by
\begin{equation}
    \bar{f}(t) =
    \frac{f(t) - \langle f \rangle
    }{
    \sqrt{
    \int
    \left( 
    f(t) - \langle f \rangle
    \right)^2
    \, {\rm d}t
    }
    }
\end{equation}
where $\langle f \rangle$ is the average of $f(t)$ over the 750 Myr of evolution considered.

For {\small LYRA}, we find (as shown in Fig. \ref{fig:temporal}) a $c$ value of $0.671$ for low-density SNe, $0.287$ for high-density SNe, and $0.661$ for all SNe, showing that the energy injection by low-density SNe are far more correlated to the energy flux than those by the high-density SNe, and still more correlated than those by all SNe combined. Similar behavior is seen for {\small RIGEL} at heights\footnote{For a comparison with {\small LYRA} at the same height of 1 kpc, the temporal correlations are: 0.387 for low-density SNe, 0.167 for high-density SNe, 0.349 for all SNe. This is a weaker overall correlation between SNe and outflows, but low-density SNe maintain a notably stronger correlation than high-density SNe, consistent with the behavior at other heights.} closer to the disk ($\pm 0.25$ instead of $\pm 1$ kpc), with $c=0.650$ for low-density SNe and $c=0.491$ for high-density SNe. There is somewhat higher correlation between the outflow and the energy from high-density SNe than in {\small LYRA}, leading to a correlation $c=0.656$ for all SNe combined, which is slightly higher than that of the low-density SNe alone. This is expected as the measurement height is now closer to the sites of star-formation, allowing even SNe with a limited range to better influence the energy flow. Nevertheless, the high-density correlation is still noticeably lower than that of the low-density SNe, and the peaks in energy from low-density SNe are likewise aligned with the outflow history, supporting the {\small LYRA} paradigm.

\newnew{In {\small LYRA}, the $\tau$ values at which the maxima occur are $-6$ Myr for low-density SNe and $+4.5$ Myr for high-density SNe. In {\small RIGEL}, they are $0$ and $+10$ Myr respectively. The energy outflow peaks tend to occur \textit{before} the peaks from the low-density SNe themselves, consistent with a nonpositive $\tau$. A high correlation alongside a positive $\tau$ does not establish that the former curve in Eq.~\ref{eq:crosscorr} causes the behavior of the latter; accordingly, it is also unphysical to conclude from the $\tau$ values that the energy flows cause the SNe to release energy. An alternative explanation may be that later SNe within the same cluster are less efficient at driving outflows than earlier ones. Later SNe go off in hotter, more pressurized environments that are more difficult to expand in, are subject to efficient radiative cooling, and might not induce a supersonic shell due to the greater ambient sound speed.}


In both simulations, the peaks of the energy outflow history (orange curve) in Fig.~\ref{fig:temporal} reach an order-unity fraction compared to the peaks of the energy released by low-density SNe (red curve), indicating that a significant fraction of energy from these SNe manages to break above and below the disk before radiative losses begin to dominate. In the analytic Sedov-Taylor solution, radiative losses are dominant only after the energy-conserving phase ends, suggesting that the expansion of many of these SN remnants are still in this phase and thus cannot reach the momentum-conserving (or ``snowplow'' phase) until after they are beyond the disk.

\subsection{Discussion}\label{subsec:analysisdisc}

The radius at which the energy-conserving phase ends, or cooling radius, of an idealized SN is
\begin{equation}\label{eq:rcool}
    r_{\rm cool} = 22.6 \, {\rm pc} \,
    \left(\frac{E}{10^{51}\,{\rm erg}}
    \right)^{0.29}
    \left(\frac{\rho}{\mHcc}
    \right)^{-0.42} \,,
\end{equation}
via Eq.~8 in \citet{KimOstriker15}, based on homogeneous and isotropic conditions\footnote{Related work \citep[e.g.,][]{Cioffi88,Thornton98} gives similar functions under slightly different numerical assumptions on cooling function.}. Realistic galaxy conditions are neither homogeneous nor isotropic, but this formula provides an order-of-magnitude estimate for the length scale of SNe-related features nonetheless. For a local density $\rho\sim10^{-4} \, \mHcc$, belonging to the low-density mode, the cooling radius is on the kiloparsec scale, larger than the galaxy's cold disk height. As such, the energy-conserving phase of a single SN blastwave is expected to proceed beyond the disk, so that energy can efficiently break into the CGM before being subject to significant radiative losses. Contrarily, for a density of $\rho \sim 10^4 \, \mHcc$ in the high-density mode, the cooling radius is only $\sim 10^{-1}\,{\rm pc}$; enough to clear the dense gas in its immediate environment, but too limited to become an outflow.

This simple estimate is consistent with the behavior of the two SN modes described in the previous sections. The environments of most low-density SNe are energetic enough to overcome their local gravitational potential and escape as far as the CGM. \new{Their high specific energies and kinetic-to-thermal energy ratios are consistent with the energy-conserving or Sedov-Taylor phase.} Thus a large amount of energy ought to be retained as they expand to their kiloparsec-scale cooling radius, with radiative losses not yet dominant. That the energy outflow history is temporally correlated with the energy released by low-density SNe is further evidence of low-density SNe efficiently imparting energy into the CGM. Contrarily, high-density environments retain too little of their energy to escape the galactic disk. Their cooling radii are also far below the galactic scale, and the energy outflow history is not correlated with the energy released by such SNe. \new{In {\small LYRA}, their environments are dominated by kinetic energy, suggesting that they have reached the snowplow, or momentum-conserving phase, where radiative energy losses are significant before expanding to the scale of the disk height.}


Star formation, as stipulated by the density threshold $\rho_{\rm th}$, can only take place in dense regions. This is consistent with results in Table~\ref{tab:snstats} showing that almost all $\menv=20\,\msun$ {\small LYRA} regions surrounding low-density SNe have zero star formation, both before (99.7\%) and after (99.8\%) the SN. It therefore stands to reason that low-density SNe cannot play a significant role in direct SF suppression. SF in high-density environments, however, are greatly affected by SNe. 23.8\% of {\small LYRA} SNe have zero star formation in the snapshot preceding the event; that is, 76.2\% of high-density SNe had \textit{nonzero} star formation in their environments before going off. After the SN, the local star formation is almost invariably reduced to zero (zero SFR in 98.4\% of cases, or nonzero SFR in only 1.6\% of cases). As such, high-density SNe are concentrated in regions of high star formation and are very effective at quenching these environments.

It is therefore reflected by the high-resolution simulations that the two modes of SNe, low-density and high-density, are each responsible for different and distinct functions. The low-density mode is responsible for outflows, but not SF suppression; the high-density mode is responsible for SF suppression, but not outflows. On a galaxy scale, this means that these two large-scale \newnew{or} \textit{macroscopic} processes take place in entirely distinct channels even as they are both driven by SN feedback.


\begin{figure*}[t]
\centering
\includegraphics[width=0.99\textwidth]{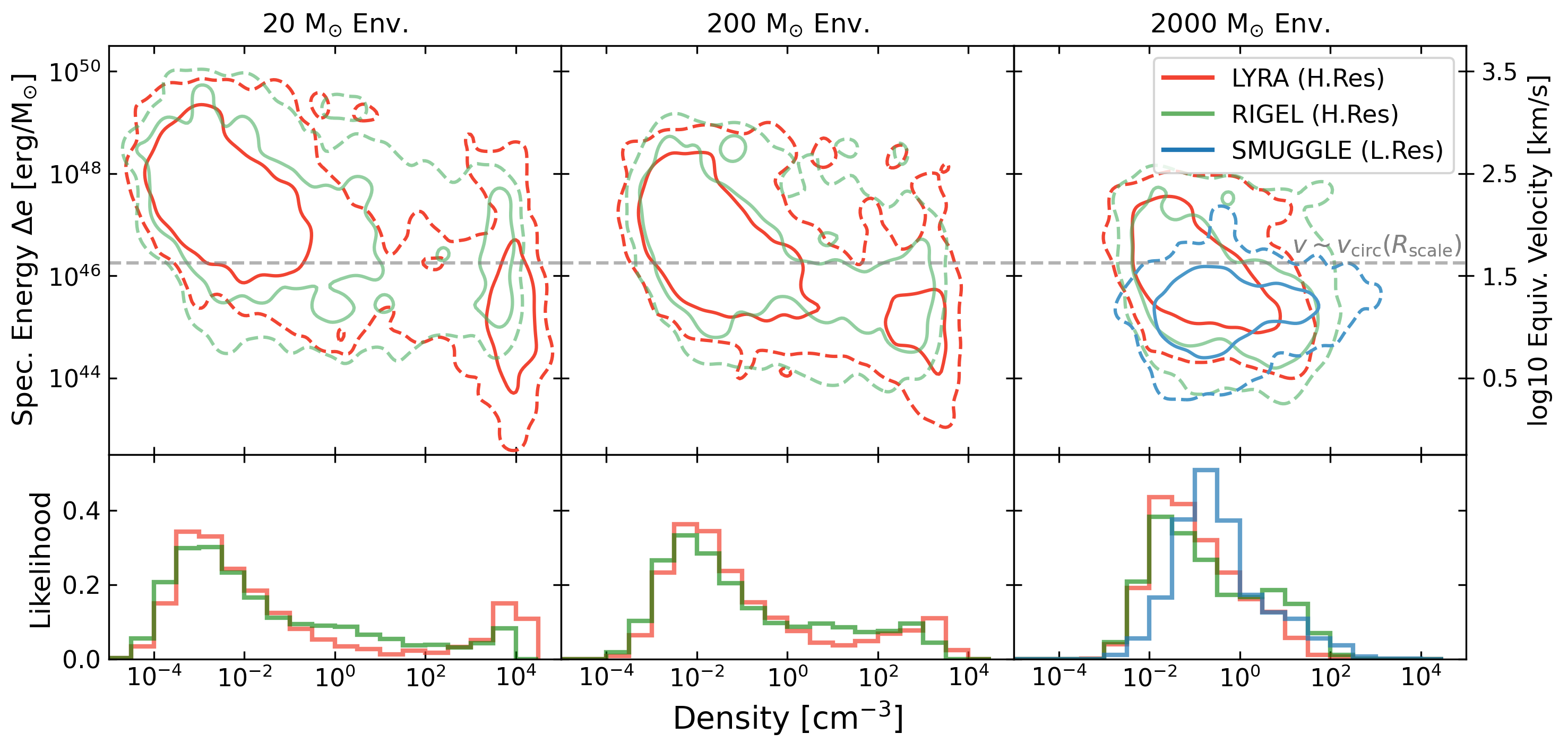}
    \caption{\textit{Impact of resolution on distinguishing the two SN modes.} $\Delta e$ as a function of $\rho$ for varying local environment size $\menv=20,\,200,\,2000\,\msun$ (\textbf{columns}), analogous to Fig.~\ref{fig:BasicEnergy}. \textbf{Top Row:} The $\Delta e$-$\rho$ scatter with the colored solid and dashed lines corresponding to 68\% and 95\% contours. The $\Delta e$ corresponding to the circular velocity (Eq.~\ref{eq:velocity}) at the gas disk scale radius is shown in gray. \textbf{Bottom Row:} Histogram of SNe densities. All LYRA results are in red, and all RIGEL results are in green. The bimodality seen in Fig.~\ref{fig:BasicEnergy} is washed out with increasing $\menv$, and is indistinguishable by $2000\,\msun$. \newnew{The lower-resolution SMUGGLE model, shown in blue, is consistent with both high-resolution simulations at the scales it can resolve.}}\label{fig:resolution}
\end{figure*}

The above paradigm is highly consistent with all results from {\small LYRA}, and in {\small RIGEL}, for the behavior of low-density SNe. For high-density {\small RIGEL} SNe, a higher proportion appear to be associated with outflow-causing behavior, e.g., a greater $\Delta e$ and a more thermal-favored ratio compared to kinetic energy. \new{According to Fig.~\ref{fig:dt}, when using a $\Delta t$ of 10 Myr for {\small LYRA}, the $\Delta e$ and kinetic-to-thermal ratio distribution among high-density SNe also appears to be the same as those of low-density SNe, matching the behavior of {\small RIGEL}. This suggests using a long $\Delta t$ can cause high-density environments to appear as if they behave like low-density environments. Indeed, for regions of equal mass $\menv$, lower-density regions occupy more volume than high-density regions, and when further considering the orbital motion of all regions over a longer $\Delta t$, it is especially common for high-density environments to overlap with low-density environments in the post-SN snapshot.}

We therefore choose to defer to the results at $\Delta t = 1\,{\rm Myr}$ where possible. Based on the data available to us, we cannot claim with certainty that a finer $\Delta t$ for {\small RIGEL} would remove all discrepancies with high-density {\small LYRA} SNe behavior. However, the weaker correlation with the energy flow in Fig.~\ref{fig:temporal} and the analytic cooling radius of $\lesssim 1\,{\rm pc}$ by Eq.~\ref{eq:rcool} do still suggest that such high-density SNe do not \newnew{meaningfully contribute to the energy within the} outflows, consistent with the overall paradigm.

\section{Impact of Resolution}\label{sec:resolution}

\subsection{High-Resolution Simulations at Coarse Mass Scales}\label{subsec:coarsening}

To study the effects of resolution on outflows and star formation, we treat the high-resolution simulations as if they were coarser, by allowing $\menv$ (as defined in Sec.~\ref{sec:environment}) to vary from fine to coarse mass scales. In this section, we use 20, 200, and 2000 $\msun$ instead of simply $M_{\rm env} = 20\,\msun$. Only the highest-resolution ``galaxy-scale'' simulations can resolve regions of $20\,\msun$, and the finest structure that most simulations are capable of modeling is typically $\gtrsim2000\,\msun$, with cosmological simulations being even coarser (e.g., $m_{\rm gas} \sim 10^4-10^6$). In this way, a number of “lumped-together” cells in these high-resolution simulations might correspond to a single mass \newnew{parcel} in a coarse-resolution simulation. The higher number of cells in our high-resolution simulation thus mimics structure within that single mass \newnew{parcel}, giving information that would not be available to a coarse-resolution simulation.\footnote{We clarify that $\menv$ is not the same as the target gas resolution, since the regions comprising $\menv$ must contain multiple gas cells to be well-sampled.}

Fig.~\ref{fig:resolution} shows a \new{contour plot} of $\Delta e$ vs.~$\rho$, analogous to Fig.~\ref{fig:BasicEnergy}, but with varying $\menv$ in each column. The bimodality distribution of densities discussed in Sec.~\ref{sec:environment} starts to fade by $\menv=200\,\msun$, and becomes \textit{completely washed out} by $\menv=2000\,\msun$, so that low-density and high-density SNe become indistinguishable. As a result, the difference between the two modes of SNe, which have significantly different physical impacts for the whole galaxy, are \textit{no longer distinguishable} when averaging over regions as large as $2000\,\msun$. In coarse-resolution simulations that can only model structure as fine as $2000 \,\msun$, low- and high-density SNe and their respective macroscopic effects thus become indistinguishable even if same the $\rho-\Delta e$ distribution at $\menv=2000\,\msun$ is maintained.

\newnew{In the rightmost panels, we additionally show results of a simulation run at such a lower resolution of $m_{\rm gas}$ = 200 $\msun$. This simulation, an analogue of the {\small LYRA} galaxy,\footnote{We use the same parameters for the initial gas and stellar disk. $\epsilon_{\rm SF}$ is set to 1.0 and $\rho_{\rm th}$ is set to $10^{-0.4}$ for a similar SF timescale to {\small LYRA} of 80 Myr, and we use a snapshot cadence of $\Delta t$ = 1 Myr.} uses a different galaxy formation model that is adapted to coarse resolutions, {\small SMUGGLE} \citep[][]{SMUGGLE}. Indeed, galaxy formation at this resolution must inherently be modeled differently because (1) it becomes impossible to resolve individual massive stars and (2) the numerical overcooling problem cannot be overcome. Therefore, the {\small SMUGGLE} model primarily differs in two ways: star particles represent aggregate stellar populations instead of single stars, and SNe inject momentum directly into the surrounding gas cells via a subgrid mechanical feedback scheme \citep[e.g.,][]{Hopkins18SNFeed, SmithSijackiShen18} instead of thermal energy. We see that under the {\small SMUGGLE} model, the $\rho-\Delta e$ distribution is indeed consistent with its high-resolution counterparts.}



As Fig.~\ref{fig:resolution} demonstrates that the density distribution is highly dependent on the choice of $\menv$, so too are individual local density estimates similarly sensitive. For a single SN, the density estimate $\rho_{2000}$ can be off by two orders of magnitude or more compared to $\rho_{20}$. This means that at coarse resolution, regions with wide differences in density are lumped together and averaged over. By the discussion in Sec.~\ref{subsec:analysisdisc}, the \new{macroscopic} behaviors resulting from SNe are well-described by estimating the cooling radius, which is a strong function of the local density $\rho$ (see Eq.~\ref{eq:rcool}). It is therefore necessary to accurately estimate $\rho$ when determining the ultimate impact of a given SN, and the most accurate estimate comes from the smallest-scale resolved structure (e.g., $\rho_{20}$ instead of $\rho_{2000}$) since Eq.~\ref{eq:rcool} is based on a homogeneous environment.
\begin{figure}
\centering
\includegraphics[width=0.98\columnwidth]{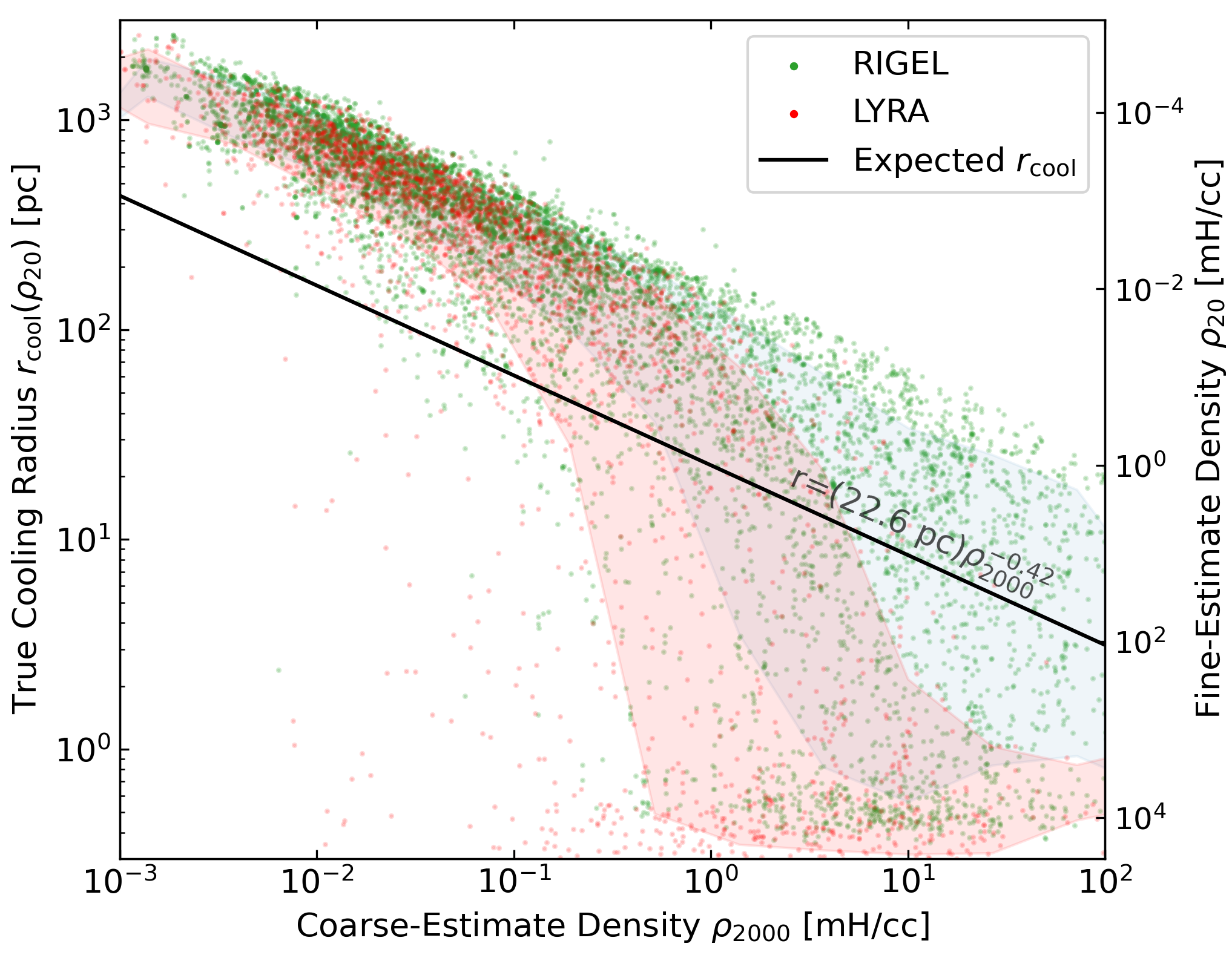}
    \caption{\textit{Mis-estimation of cooling radii in coarse resolution contexts.} The ``true'' cooling radius $r_{\rm cool}$ as a function of the coarse-estimate local density $\rho_{2000}$, for LYRA (red) and RIGEL (green). Each point represents a single SNe, and the shaded regions represent the 1-$\sigma$ range of cooling radii at each density. The true cooling radius is calculated by Eq.~(\ref{eq:rcool}) using the fine-estimate local density $\rho_{20}$. As such, it is a proxy for the fine local density, whose corresponding values are shown on the right. The solid black line shows the cooling radius calculated directly from $\rho_{2000}$ instead of $\rho_{20}$. At low $\rho_{2000}$, this line underestimates the true $r_{\rm cool}$ by as much as $5 \times$; at high $\rho_{2000}$, the true $r_{\rm cool}$ can vary by up to 3 orders of magnitude about this line, with a bimodality towards the largest and smallest possible radii.}
    \label{fig:coolingradius}
\end{figure}

We examine in Fig.~\ref{fig:coolingradius} the relation between the ``true'' cooling radius (as estimated with the fine-grained $\rho_{20}$) and the coarse-grain density $\rho_{2000}$. Since $r_{\rm cool} (\rho_{20})$ is itself a direct function of $\rho_{20}$ (when assuming an energy of $10^{51}$ erg), this can also be understood as a relation between the fine- and coarse-grain densities. For comparison, we also plot the direct cooling radius vs. density relationship as a function of $\rho_{2000}$ as a solid black line. This line represents the expected length-range when only coarse-resolution information (such as $\rho_{2000}$) is available, whereas the points correspond to the expected length-range of a SN's effect when higher-resolution information (such as $\rho_{20}$) is available. 

Recalling from Sec. \ref{subsec:analysisdisc} that a large value of $r_{\rm cool} (\rho_{20})$ \newnew{indicates a correlation with large-scale outflows}, and a small value corresponds to a tendency to locally suppress star formation, Fig.~\ref{fig:coolingradius} thus expresses the extent to which the coarser estimate $\rho_{2000}$ fails to predict the \new{true resulting macroscopic behavior} of SNe. In particular, for low $\rho_{2000}$, the points tend to lie above the solid line by as much as a factor of 5. This means that estimating the local ISM density based on the coarse structure would typically \textit{underestimate} the SN’s true cooling radius, predicting shorter-ranged SNe effects than are accurate. Meanwhile, at higher $\rho_{2000}$, a given $\rho_{2000}$ value can correspond to a wide range of cooling radii. As such, the behavior of the SN is not correlated with the coarse-average ISM density at all, and there is instead a bimodality towards the largest and smallest radii within this range. These results therefore suggest caution when using coarse estimates of density to determine $r_{\rm cool}$, e.g., in subgrid models where the cooling radius is used to designate the energy-conserving phase of an SN as numerically unresolved.

Consider, for example, a single SN going off in a coarse-resolution density (i.e., $\rho_{2000}$) of $1\,\mHcc$. If assumed to go off in a homogeneous 2000 $\msun$ region (as it likely would, if the region was sampled by only $\mathcal{O}(1)$ cells), it would have a cooling radius of $\sim20\,{\rm pc}$. Beyond this radius, it would transition to the snowplow phase and carry most of its energy as momentum. In a coarse-resolution simulation \newnew{(e.g., {\small SMUGGLE})} that cannot spatially resolve this cooling radius, direct injection of thermal energy leaves the dynamics susceptible to the numerical overcooling problem. \newnew{In {\small SMUGGLE}, the SN is thus initialized by directly injecting momentum into the surrounding gas cells isotropically.}


However, we have established, this 2000 $\msun$ region is almost certainly not homogeneous, and a $\rho_{2000}$ of $1\,\mHcc$ might correspond to a wide range of fine-resolution densities $\rho_{20}$ and thus a wide range of cooling radii. According to Fig.~\ref{fig:coolingradius}, this same SN might have a true cooling radius as low as $\sim0.5\,{\rm pc}$ or as high as $\sim300\,{\rm pc}$. In the former case, it would kinetically disrupt the immediately surrounding dense ($\sim10^4\,\mHcc$) star-forming cloud, but at a distance scale far smaller than a typical coarse-resolution gas cell. In the latter case, it would reach the end of the Sedov-Taylor phase at a distance comparable to the scale height of the whole galaxy (0.35 kpc), imparting energy into a highly inhomogeneous region reaching above and below the disk. Neither of these outcomes is consistent with an isotropic momentum injection at $\sim 20\,{\rm pc}$, as would be the result for a homogeneous region.

\begin{figure}[t]
\centering
\includegraphics[width=0.95\columnwidth]{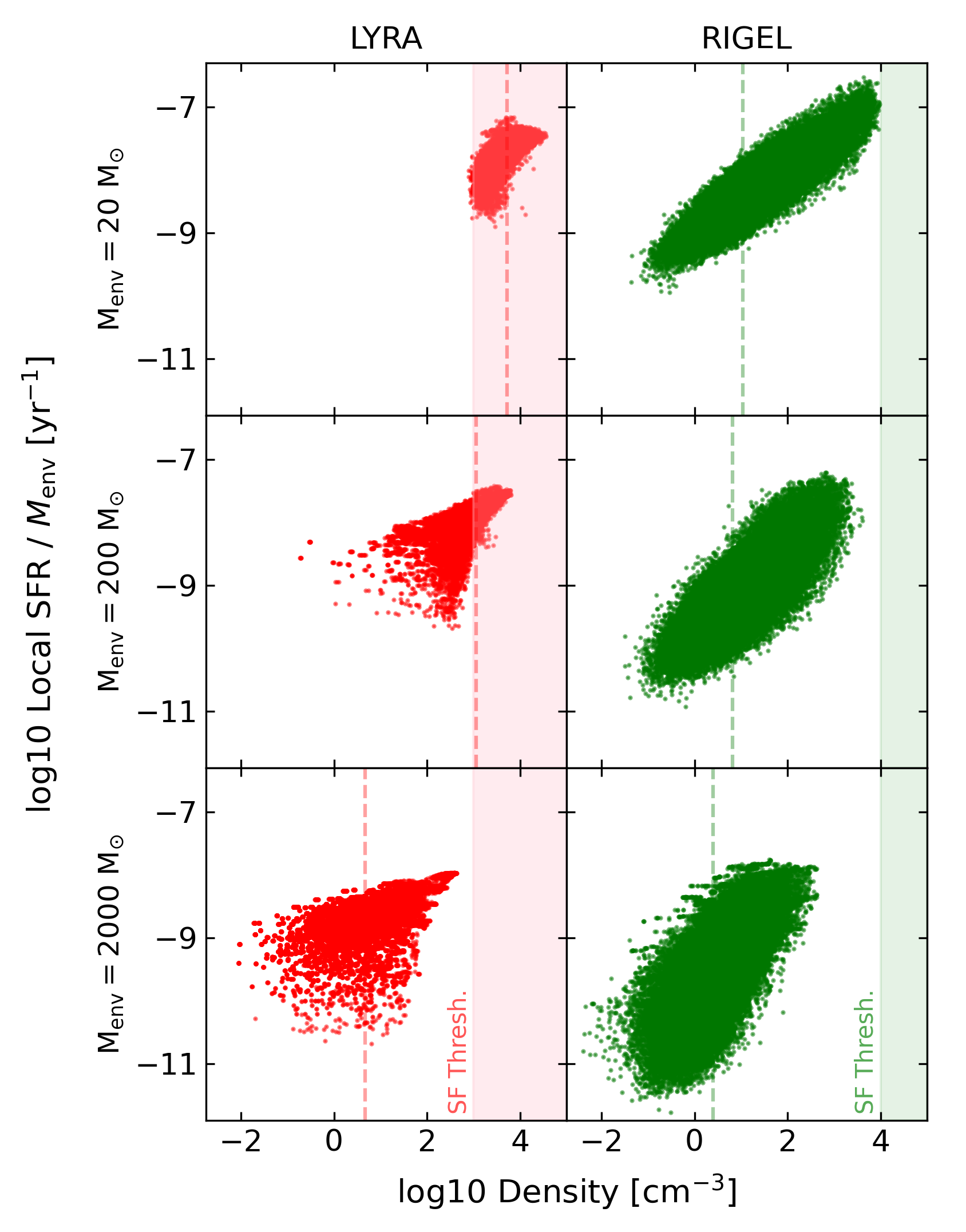}
    \caption{\textit{Properties of star-forming regions.} SFR of star-forming regions versus the local density, for varying $\menv$ (rows) for LYRA (left column) and RIGEL (right column), across snapshots every 10 Myr. The densities $\rho_{\rm th}$ at which individual gas cells are eligible for star formation in each simulation are shaded. The median density of all star-forming regions of mass $\menv$ is shown in each panel as a vertical dashed line. For coarse $\menv$, it is common for regions far below the per-cell SF threshold to still be star-forming. At $\menv = 2000\,\msun$, even regions with densities as low as $10^{-2}$ $\mHcc$ have nonzero star formation rates.}
    \label{fig:sfregions}
\end{figure}

\subsection{SF Suppression}\label{subsec:res_sf_supp}

The same coarsening process described in \ref{subsec:coarsening} may also be applied to star-forming \textit{gas} cells instead of star particles. Instead of the $\Delta e$ surrounding star particles, we now calculate the total star formation rate of the nearest $\menv$ for each SF-eligible gas cell. We remind the reader that the total SFR of a region is the sum of the cell-wise SFR’s of its eligible gas cells, as defined by Eq.~\ref{eq:cellSFR}, and that gas cells are considered eligible if they fulfill all the requirements for star formation outlined in Sec.~\ref{subsec:sfrmodel} for their respective simulations. We then plot this local SFR against the density $\rho$ for $\menv=20,200,2000 \, \msun$. The results are shown in Fig.~\ref{fig:sfregions}, where cells from snapshots every 10 Myr are included.

{\small LYRA} and {\small RIGEL} have cell-wise SF thresholds $\rho_{\rm th}$ of $10^3$ and $10^4$ $\mHcc$ respectively for individual cells, but Fig.~\ref{fig:sfregions} shows that larger regions need not have such high densities to have nonzero star formation. When considering regions as large as $\menv=2000\,\msun$, the median star-forming region has a density of $\sim 10^{0.5} \, \mHcc$. Even volumes with densities as low as $10^{-2}$ to $10^{-1}$ $\mHcc$ can still have nonzero star formation rates, suggesting that using a high $\rho_{\rm th}$ at such coarse resolutions can \textit{artificially} prevent regions from forming stars.

It is still true at larger mass scales that low-density SNe do not suppress star formation directly, whereas high-density SNe do. Table~\ref{tab:ressfstats} shows statistics on the proportion of (non) star-forming regions before and after the event, for both modes of SNe in {\small LYRA}\footnote{As discussed in Sec.~\ref{subsec:indivSNe}, the snapshot interval of {\small RIGEL} is not frequent enough to resolve rapid changes in the SFR as clumps are disrupted.}. Crucially, we distinguish low- and high-density SNe based on the \textit{fine-grained} $\menv = 20 \,\msun$ information from Sec.~\ref{sec:environment}, as we have established in this section that coarse-grained density estimates cannot reliably produce this distinction. We additionally present, for each SN mode, the fraction of \textit{star-forming} regions that later become quenched; that is, the number of regions with nonzero SFR before the SN \textit{and} zero SFR afterward, divided by all regions with nonzero SFR before. We call this the \textit{quenched fraction} and quote its values for each mode in Table~\ref{tab:ressfstats} as well.

For low-density SNe, the proportion of zero-SFR regions is similar before and after the SN for all $\menv$, indicating a negligible impact on SF suppression by these SNe at all scales. This is consistent with the paradigm established in Sec.~3.3. High-density SNe are still very effective at suppressing SF, but not with totality for wider regions (e.g., 74\% for $\menv=2000\,\msun$). The quenching fractions reflect this as well, where low- (high-) density SNe have low (high) quenching rates. The exception to this is low-density SNe at $M_{\rm env}=20\,\msun$, where 6/6 SNe quench their surroundings; but this applies to only a very small number of all SNe.

\begin{deluxetable}{lrrr}
\tabletypesize{\footnotesize}
\tablecaption{SF Region Statistics for LYRA SNe}
\tablehead{
Quantity & $\menv=20\,\msun$ & $\menv=200\,\msun$ & $\menv=2000\,\msun$
}
\startdata
\textbf{Low Dens.} & & & \\
SFR=0 Bef. & 99.7\% & 88.6\% & 72.7\% \\
    SFR=0 Aft. & 99.8\% & 87.9\% & 73.0\% \\
    Quenched&6 / 6 (100\%)& 59 / 250 (24\%)& 71 / 596 (12\%)\\
    \\
    \textbf{High Dens.} & & & \\
    SFR=0 Bef. & 23.8\% & 19.4\% & 15.0\% \\
    SFR=0 Aft. & 98.4\% & 88.3\% & 74.0\% \\
    Quenched&408 / 416 (98\%)& 384 / 440 (87\%)& 328 / 464 (71\%)
\enddata \label{tab:ressfstats}
\tablecomments{The first two rows for each mode are analogous to the last two columns of Table~\ref{tab:snstats}, but for varying $\menv$. The quenched fractions compare the number of regions that are star-forming before the SN but not after the SN, to the total number of regions that are star-forming beforehand.}
\end{deluxetable}

\section{Implications for Galaxy Simulations}\label{sec:discussion}

\subsection{Subgrid Models for SNe}\label{subsec:discussion_subgrid}

By Fig.~\ref{fig:resolution}, the transition of $\menv$ at which differences between the two SN modes, and thus \new{their resulting macroscopic effect} on the galaxy, can be distinguished is between 200 and 2000 $\msun$. For a simulation seeking to model such differences, these environments must contain multiple \newnew{gas cells} to be resolved; typically on the order of 10 cells. This would suggest that the threshold for the necessary mass per cell lies between $\sim 20$ and $\sim 200$ $\msun$ per cell; that is, the mass resolution should be $\lesssim 10^2\,\msun$.

\new{Simulations with better resolution than this can distinguish between these modes naturally; however, for mass-discretized codes in which the smallest unit of resolvable mass is $m_{\rm gas} \gg 10^2 \, \msun$, \textit{effective} models, perhaps informed by high-resolution results, may be used to account for inhomogeneous sub-grid densities. While more accurate, such models would not be self-consistent because they depend on information not directly contained within the hydrodynamics solver.}

One popular approach in the literature addressing this aspect is to decouple outflows from the fluid mesh, a method used by well-established volume-simulations like Illustris or TNG, or {\small ARKENSTONE} \citep[e.g.,][]{Vogelsberger2014,Pillepich18TNG,ARKENSTONE1}. These ``decoupled wind'' models may address this challenge particularly well, since they (1) do not \textit{explicitly} depend on local ISM conditions to deposit and retain feedback, and (2) inherently restrict outflow energy to low-density gas as winds are only set to hydrodynamically recouple below some density threshold (e.g., $10^{-2} \, \mHcc$). Bipolar feedback models \citep[e.g.,][]{Vogelsberger13,Zhang24} may also mimic the large-scale effects of SNe well, though they still rely explicitly on external conditions such as the orientation of the disk, assuming a well-formed one exists.

Fluid solvers using adaptive mesh refinement (AMR), such as those used in e.g., \citet{Walch15SILCC,Agertz20EDGE}, will likely suffer less from this problem, as they can reach much finer mass discretization where needed, ensuring that feedback is coupled to the appropriate amount of mass and allowing resulting behaviors to emerge naturally. Even so, fully resolving star-forming regions (i.e., the Jeans length of dense clumps) with AMR solvers is computationally expensive, and the outflow properties are still sensitive to numerical parameters such as the star formation efficiency $\epsilon_{\rm SF}$ \citep[e.g.,][]{Hu23Solvers}. Similar strategies, such as super-Lagrangian refinement near shock fronts \citep[commonly used in AGN treatments, e.g.,][]{Sivasankaran22}, may also allow the necessary small-scale structure to be self-consistently traced, though they too may become expensive for galaxies with a large number of stars.

Ultimately, the difficulty of \new{correctly capturing the macroscopic effects} of SNe are rooted in the \new{numerical} overcooling problem. If high-temperature gas could be properly handled by the fluid solver (as in high-resolution cases), the preferred solution of direct thermal energy injection would again be feasible. The hottest ($\sim 10^6$ K) ISM gas fills up a large volume fraction despite taking up only a low mass fraction. As a result, it becomes diluted among the more massive warm and cold phases in lower-resolution simulations, even as it is subject to drastically different cooling physics \citep[e.g.,][]{Wiersma09}. To that end, subgrid models of the gas itself, accounting for separate hot and cool phases within the gas \newnew{cells} using a multi-fluid approach \citep[e.g.,][]{Scannapieco06,Weinberger23,Das24MOGLI} may be a promising avenue of exploration. 

\subsection{Star-Forming Regions}

Models aiming to explicitly resolve the multi-phase nature of gas tend to restrict star formation to only higher density regions, so as to capture the conditions in molecular clouds. Our results in Fig.~\ref{fig:sfregions} showing nonzero SFR in 2000 $\msun$ regions with densities down to $10^{-2}\,\mHcc$ suggest that it may not be physically correct to impose a very high density threshold $\rho_{\rm th}$ for star formation in coarse resolution simulations. Lower star formation thresholds are used in some cosmological-volume simulations that model gas with an effective equation of state \citep[e.g.,][]{Springel2003,Vogelsberger13,Vogelsberger2014,Schaye15EAGLE,Pillepich18TNG}. The value of $\rho_{\rm th}$ is ultimately an important question to address as it has been shown to correlate with the level of burstiness, outflow generation, and dark matter core formation in some dwarf simulations \citep[e.g.,][]{PontzenGovernato12,Dutton2019}.

For example, in Fig.~\ref{fig:sfregions}, a region consisting of 2000 $\msun$ may possess a nonzero star formation rate at average densities of $\rho \sim 0.1$ or even $\sim 0.01$ $\mHcc$. We emphasize that this is true even as {\small LYRA} and {\small RIGEL} have their SF thresholds set at the much higher levels of $10^3$ and $10^4$ $\mHcc$, respectively, when resolving $\sim {\rm M}_\odot$ scales.

\subsection{Caveats}\label{subsec:caveats}

We have posited a dichotomy between outflows and the \textit{direct} suppression of star formation via SNe. This does not preclude \textit{indirect} suppression via outflows; outflows are still capable of removing total mass from the galaxy, which may still lead to eventual quenching by depleting the total available gas supply on longer timescales \citep[$\sim {\rm Gyrs}$, e.g.,][]{Bigiel08,Bigiel11,Schiminovich10,Leroy13}. Efficient deposition of energy from outflows to the CGM may also be responsible for \textit{preventative} feedback \citep[e.g.,][]{Carr23,Pandya23,Bennett24,Voit24,Wright24}, which slows star formation not by disrupting star-forming regions but by preventing gas from accreting onto the disk in the first place. In this manner, energetic outflows do cause a lower overall SFR, but only by affecting gas that is not yet within the galaxy. This mechanism is not precluded by the dichotomy; contrarily, scenarios where outflows directly remove SF-eligible gas from the disk (e.g., types of \textit{ejective} feedback) may be disfavored.

Feedback from stellar radiation is not included in the simulations studied. While this makes it easier to directly determine effects from SNe, stellar radiation is known to have significant impacts on the galaxy, in particular in the early dispersal of star forming clumps and the reduction in supernova clustering. In the full {\small RIGEL} simulation including radiative transfer \citep[presented in][]{Deng24RIGEL}, the high-density mode of SNe entirely disappears so that radiative feedback completely eclipses the role of SNe in SF suppression. On the scale of dwarf galaxies, these effects have also been previously reported \citep[e.g.,][]{VazquezSemadeni10,Walch12,Dale14,Sales14,Rosdahl15PI,Emerick18Rad}, but the full disappearance of the high-density mode due to radiation is not ubiquitous, as the efficiency of radiative feedback is expected to saturate under sufficiently high surface densities \citep[e.g.,][]{Dekel23}.

\section{Conclusions}\label{sec:conclusion}

We analyze two ultra-high resolution simulations of similar dwarf galaxies, {\small LYRA} and {\small RIGEL}, which have target mass resolutions 4 $\msun$ and 1 $\msun$, respectively. Within the context of these full galaxy simulations, we focus specifically on the role of SNe on their local environments, and their resulting effect on driving galactic outflows and suppressing star formation. Our main findings are as follows:

\begin{enumerate}

    \item Consistent with other high-resolution studies, the densities in which SNe go off follow a bimodal distribution: a low-density mode, and a high-density mode.
    
    \item Each of these modes is primarily \newnew{correlated with} a different \textit{macroscopic}, galaxy-scale process. The low-density mode \newnew{is mostly linked to galactic outflows carrying energy into the CGM, and the} high-density mode tends to suppress star formation via direct dispersal of dense star-forming clumps.
    
    \item The two modes of SNe can only be distinguished by their densities on \textit{fine} mass scales ($m_{\rm cell} \lesssim 10^2\,\msun$). On coarser $m_{\rm cell} \gg 10^2\,\msun$ mass scales, highly inhomogeneous regions are averaged over, causing the two SN modes and their resulting \new{macroscopic} behaviors to become indistinguishable. \new{}

    \item The true cooling radius of SNe cannot be reliably determined from the density on coarse scales, affecting decisions in subgrid models (e.g., whether a SN should be initialized as thermal energy or as momentum). \new{As a result, subgrid models in coarse-resolution simulations cannot accurately trace these behaviors in a \textit{self-consistent} manner.}

    \item When coarse mass scales are averaged over, even diffuse regions with densities down to $10^{-2} \, \mHcc$ can have nonzero star formation rates. This is a significantly lower threshold than those used to mimic dense molecular clouds at high resolution.

\end{enumerate}

In light of these findings, we conclude that simulations seeking to model the way SNe drive outflows and suppress star formation in a self-consistent fashion must have a high requisite resolution of $m_{\rm cell} \lesssim 10^2\,\msun$. Alternatively, coarse-resolution simulations may model SNe with fidelity to high-resolution results, via effective models that mimic their emergent effects.

\section*{Acknowledgments}

The authors thank Thorsten Naab for helpful discussions which helped improve the first version of this draft, and for contributions to the LYRA simulation data from which this work draws upon. EZ further thanks Jose Benavides and Simeon Bird for providing additional clarifying insights on the ideas presented in this work. EZ and LVS acknowledge the financial support received through NSF-CAREER-1945310, NSF-AST-2107993, and NSF-AST-2408339 grants. Computations were performed using the computer clusters and data storage resources of the HPCC, which were funded by grants from NSF (MRI-2215705, MRI-1429826) and NIH (1S10OD016290-01A1). TAG acknowledges the computing time provided by the Leibniz Rechenzentrum (LRZ) of the Bayerische Akademie der Wissenschaften on the machine SuperMUC-NG (pn73we). HL is supported by the National Key R\&D Program of China No. 2023YFB3002502, the National Natural Science Foundation of China under No. 12373006 and 12533004, and the China Manned Space Program with grant No. CMS-CSST-2025-A10. FM gratefully acknowledges funding from the European Union – NextGenerationEU, in the framework of the HPC project – “National Centre for HPC, Big Data and Quantum Computing” (PNRR – M4C2 – I1.4 – CN00000013 -– CUP J33C22001170001).



\bibliographystyle{aasjournal}
\bibliography{biblio}

\label{lastpage}

\end{document}